%% file: vMEM-arXiv.tex
\documentclass[a4paper,oneside,notitlepage,onecolumn,12pt]{article}
\usepackage[paper=a4paper, left = 3cm, right = 3cm, top = 3cm, bottom = 3cm]{geometry}
\usepackage[latin1]{inputenc}
\usepackage{amssymb,amsmath}
\usepackage{dsfont}
\usepackage{graphicx}
\usepackage[authoryear,round]{natbib}
\usepackage{rotating}
\usepackage{setspace}
\usepackage{url}
\usepackage{bm}
\usepackage{multirow}
\usepackage{times}
\usepackage[usenames,dvipsnames]{color}
\usepackage{graphicx}
\usepackage{caption}
\usepackage{subcaption}

\DeclareMathOperator{\diag}{diag}

\DeclareMathOperator{\trace}{trace}

\newcommand{\Der}[2]{\nabla_{#1}{#2}}

\newcommand{\dlim}{\stackrel{d}{\rightarrow}}

\newcommand{\R}{\mathds{R}}

\begin{document}

\title{
{Copula--based Specification of vector MEMs}
\thanks{Some of the material presented here circulated in the NBER WP 12690
(2006) ``Vector Multiplicative Error Models: Representation and Inference''.
We thank Nikolaus Hautsch, Loriano Mancini, Joseph Noss, Andrew Patton, Kevin Sheppard, David Veredas as well as participants in seminars at CORE, Humboldt Universit\"at zu Berlin, Imperial College, IGIER--Bocconi for helpful comments.
The usual disclaimer applies.
Financial support from the Italian MIUR under grant PRIN MISURA is gratefully acknowledged.}}
\author{Fabrizio Cipollini\thanks{Dipartimento  di Statistica, Informatica, Applicazioni ``G. Parenti'',
Universit\`a di Firenze, Italy. e-mail: \tt{cipollini@disia.unifi.it}} \
\quad Robert F. Engle\thanks{Department of Finance, Stern School of
Business, New York University; email \tt{rengle@stern.nyu.edu}} \
\quad Giampiero M. Gallo\thanks{Corresponding author: Dipartimento
di Statistica, Informatica, Applicazioni ``G. Parenti'', Universit\`a di Firenze, Viale G.B.
Morgagni, 59 - 50134 Firenze -- Italy. e-mail: \tt
{gallog@disia.unifi.it}}}

\date{}

\maketitle

\begin{abstract}
  The Multiplicative Error Model~(\citet{Engle:2002}) for nonnegative valued processes is specified as the product of a (conditionally autoregressive) scale factor and an innovation process with nonnegative support.
A multivariate extension allows for the innovations to be contemporaneously correlated. We overcome the lack of sufficiently flexible probability density functions for such processes by suggesting a copula function approach to estimate the parameters of the scale factors and of the correlations of the innovation processes. We illustrate this vector MEM with an application to the interactions between realized volatility, volume and the number of trades. We show that significantly superior realized volatility forecasts are delivered in the presence of other trading activity indicators and contemporaneous correlations.
\end{abstract}

\textbf{Keywords:} GARCH; MEM;  Realized Volatility; Trading Volume; Trading Activity; Copula; Volatility Forecasting.


\newpage


\setlength{\parskip}{6pt}

\section{Introduction}

Dynamics in financial markets can be characterized by many indicators of trading activity such as absolute
returns, high-low range, number of trades in a certain interval (possibly labeled as buys or sells),
volume, high--low range, ultra-high frequency based measures of volatility, financial durations and so on.

\citet{Engle:2002} reckons that one striking regularity
of financial time series is that persistence and clustering
characterizes the evolution of such processes. As a result, the
dynamics of such variables can be specified as the product of a
conditionally deterministic scale factor which evolves according to
a GARCH--type equation and an innovation term which is iid~with
unit mean. Such models are labeled Multiplicative Error Models (MEM) and can be seen as a generalization of the GARCH
(\citet{Bollerslev:1986}) and ACD (\citet{Engle:Russell:1998}) approaches. One of the advantages of such a model is to avoid the need to resort to logs (not possible when
zeros are present in the data) and to provide conditional
expectations of the variables of interest directly (rather than
expectations of the logs). Empirical results show a good performance
of these types of models in capturing the stylized facts of the
observed series (e.g. for daily range,~\cite{Chou:2005}; for
duration, volume and volatility~\citet{Manganelli:2005}; for volatility, volume and trading intensity~\citet{Hautsch:2008}).

The model can be specified in a multivariate context (vector MEM or vMEM) allowing just the lagged values of each variable of interest to affect the conditional expectation of the other variables beside its own. Such a specification lends itself to producing multi--step ahead forecasts: for example, \citet{Engle:Gallo:2006} specify a multivariate MEM where the dynamics of three different measures of volatility, namely absolute returns, daily range and realized volatility, influence each other temporally, and evaluate the contribution of MEM-based forecasts to the prediction of VIX. Although equation--by--equation estimation ensures consistency of the estimators in a quasi-maxim\-um likelihood context, given the stationarity conditions discussed by~\citet{Engle:2002}, correlation among the innovation terms is not taken into account and leads to a loss in efficiency.

The specification of a multivariate distribution of the innovations is far from trivial: joint probability distributions for nonnegative--valued random variables are not available except in very special cases. In this paper, we suggest a maximum likelihood estimation strategy adopting copula functions to link together marginal probability density functions for individual innovations specified as Gamma as in~\citet{Engle:Gallo:2006} or as zero--augmented distributions as in~\citet{Hautsch:Malec:Schienle:2014} distinguishing between the zero occurrences and the strictly positive realizations. Copula functions are used in a Multiplicative Error framework but in a Dynamic Conditional Correlation context by~\citet{Bodnar:Hautsch:2016}. As an alternative for the vector MEM, \citet{Cipollini:Engle:Gallo:2013} suggest a semiparametric approach resulting in a GMM estimator.

The range of potential applications is quite wide: dynamic interactions among different values of volatility, volatility spillovers across markets (allowing multivariate-multi-step ahead forecasts and impulse response functions, order execution dynamics (\citet{Noss:2007} specifies a MEM for execution depths). As an illustration we will concentrate on the interaction of various measures of market activity (volatility, volume and number of trades) in which the conditional expectations depend just on the past values (not also on some contemporary information as in \citet{Manganelli:2005} and \citet{Hautsch:2008}).

What the reader should expect is the following: in Section \ref{sect:MEM} we lay out the specification of a vector Multiplicative Error Model, discussing the issues arising from the adoption of several types of copula functions linking univariate Gamma marginal distributions.
In Section \ref{sect:MLInference} we describe the Maximum Likelihood procedure leading to the inference on the parameters. Section \ref{sect:TradingActivity} presents the application of our model to three series of trading activity, namely realized kernel volatility, traded volumes and the number of trades. The illustration is performed on the JNJ stock over a period between 2007 and 2013. What we find is that specifying the joint distribution of the innovations allowing for contemporaneous correlation dramatically improves the log--likelihood over an independent (i.e. equation--by--equation) approach. Richer specifications (where simultaneous estimation is unavoidable) deliver a better fit, improved serial correlation diagnostics, and a better performance in out--of--sample forecasting. The Student--T copula possesses better features than the Normal copula. Overall, the indication is that we will have significantly superior realized volatility forecasts when other trading activity indicators and contemporaneous correlations are considered.  Concluding remarks follow.

\section{Multiplicative Error Models}
\label{sect:MEM}

Let $\bm{x}_t$ be a $K$--dimensional process with non--negative components.
A vector Multiplicative Error Model (vMEM) for $\bm{x}_t$ is defined as
\begin{equation}
  \label{eqn:vMEM_mu_eps}
  \bm{x}_t = \bm{\mu}_t \odot \bm{\varepsilon}_t = \diag(\bm{\mu}_t) \bm{\varepsilon}_t,
\end{equation}
where $\odot$ indicates the Hadamard (element--by--element) product and $\diag$ is a diagonal matrix with its vector argument on the main diagonal.
Conditionally upon the information set $\mathcal{F}_{t-1}$, a fairly general specification for $\bm{\mu}_t$ is
\begin{equation} \label{eqn:muVector2}
    \bm{\mu}_t = \bm{\omega} + \bm{\alpha} \bm{x}_{t-1} + \bm{\gamma} \bm{x}_{t-1}^{(-)} +  \bm{\beta} \bm{\mu}_{t-1},
\end{equation}
where $\bm{\omega}$ is $(K,1)$ and $\bm{\alpha}$, $\bm{\gamma}$ and $\bm{\beta}$ are $(K,K)$.
The vector $\bm{x}_t^{(-)}$ has a generic element $x_{t,i}$ multiplied by a function related to a signed variable, be it a positive or negative return ($0,1$ values) or a signed trade (buy or sell $1,-1$ values), as to capture asymmetric effects.
Let the parameters relevant for $\bm{\mu}_{t}$ be collected in a vector $\bm{\theta}$.

Conditions for stationarity of $\bm{\mu}_t$ are a simple generalization of those of the univariate case (e.g.~\citet{Bollerslev:1986}; \citet{Hamilton:1994}): a vMEM(1,1) with $\bm{\mu}_t$ defined as in equation (\ref{eqn:muVector2})
is stationary in mean if all characteristic roots of
$\bm{A} = \bm{\alpha} + \bm{\beta} + \bm{\gamma}/2$
are smaller than 1 in modulus. We can think of $\bm{A}$ as the \emph{impact matrix} in the expression
\begin{equation*}
  E(\bm{x}_{t+1}|\mathcal{F}_{t-1})=  \bm{\mu}_{t+1|t-1} = \bm{\omega} + \bm{A} \bm{\mu}_{t|t-1}.
\end{equation*}

If more lags are considered, the model is
\begin{equation}
  \label{eqn:muVectorL}
  \bm{\mu}_t = \bm{\omega} + \sum_{l=1}^{L}\left[ \bm{\alpha}_l \bm{x}_{t-l} + \bm{\gamma}_l \bm{x}_{t-l}^{(-)} +  \bm{\beta}_l \bm{\mu}_{t-l}\right],
\end{equation}
where $L$ is the maximum lag occurring in the dynamics.
It is often convenient to represent the system~(\ref{eqn:muVectorL}) in its equivalent \textit{companion form}
\begin{equation}
  \label{eqn:impact_matrix}
  \bm{\mu}_{t+L|t-1}^{*} = \bm{A}^{*} \bm{\mu}_{t+L-1|t-1}^*,
\end{equation}
where $\bm{\mu}_{t+L|t-1}^* = (\bm{\mu}_{t+L|t-1}; \bm{\mu}_{t+L-1|t-1}; \ldots;  \bm{\mu}_{t+1|t-1})$ is a $KL\times 1$ vector obtained by stacking its elements
columnwise and
\begin{equation*}
  \bm{A}^{*} =
  \left(
    \begin{array}{ccccc}
       \bm{A}_1 & \bm{A}_2        & \cdots & \bm{A}_L        \\
                & \bm{I}_{K(L-1)} &        & \bm{0}_{K(L-1),K}
    \end{array}
  \right)
\end{equation*}
with $\bm{A}_l = \bm{\alpha}_l + \bm{\beta}_l + \bm{\gamma}_l/2, \ l=1,\ldots,L$, $\bm{I}_{K(L-1)}$ is a $K(L-1)\times K(L-1)$ identity matrix and $\bm{0}_{K(L-1),K}$ is a $K(L-1)\times K$ matrix of zeros.
The same stationarity condition holds in terms of eigenvalues of $\bm{A}^{*}$.


The innovation vector $\bm{\varepsilon}_t$ is a $K$--dimensional iid process with density function defined over a $[0,+\infty)^K$ support, the unit vector
$\mathds{1}$ as expectation and a general variance--covariance matrix $\bm{\Sigma}$,
\begin{equation}
  \label{eqn:veps_general}
  \bm{\varepsilon}_t|\mathcal{F}_{t-1} \sim  D^{+}(\mathds{1},\bm{\Sigma}).
\end{equation}
The previous conditions guarantee that
\begin{align}
    E(\bm{x}_t|\mathcal{F}_{t-1}) & = \bm{\mu}_t                  \label{eqn:Ex_t_general} \\
    V(\bm{x}_t|\mathcal{F}_{t-1}) & = \bm{\mu}_t\bm{\mu}_t' \odot \bm{\Sigma} = \diag(\bm{\mu}_t) \bm{\Sigma} \diag(\bm{\mu}_t), \label{eqn:Vx_t_general}
\end{align}
where the latter is a positive definite matrix by construction.

Some alternatives can be considered about the specification of the distribution of the error term ${\bm
\varepsilon}_t|\mathcal{F}_{t-1}$.

\subsection{Multivariate Gamma Formulations}
\label{sect:MultivariateGamma}

The generalization of the univariate gamma adopted by \citet{Engle:Gallo:2006} to a multivariate counterpart is frustrated by the limitations of the multivariate Gamma distributions available in the literature (all references below come from~\citet[chapter~48]{Johnson:Kotz:Balakrishnan:2000}): many of them are bivariate formulations, not sufficiently general for our purposes; others are defined via the joint characteristic function, so that they require
tedious numerical inversion formulas to find their probability density functions (pdf).
The formulation that is closest to our needs (it provides all univariate marginal probability functions for $\varepsilon_{i,t}$ as $Gamma(\phi_i,\phi_i)$), is a particular version of the multivariate Gamma's by Cheriyan and Ramabhadran (henceforth $GammaCR$, which is equivalent to other versions by Kowalckzyk and Trycha and by Mathai and Moschopoulos):
\begin{equation*}
  \bm{\varepsilon}_t|\mathcal{F}_{t-1} \sim GammaCR(\phi_0,\bm{\phi},\bm{\phi}),
\end{equation*}
where $\bm{\phi} = (\phi_1; \ldots; \phi_K)$ and $0 < \phi_0 < \min(\phi_1,\ldots, \phi_K)$ (\citet[454--470]{Johnson:Kotz:Balakrishnan:2000}).
The multivariate pdf is expressed in terms of a cumbersome integral and the conditional correlations matrix of $\bm{\varepsilon}_t$ has generic element
\begin{equation*}
  \rho(\varepsilon_{t,i}, \varepsilon_{t,j}|\mathcal{F}_{t-1}) = \frac{\phi_0}{\sqrt{\phi_{i}\phi_{j}}},
\end{equation*}
which is restricted to be positive and is strictly related to the variances $1/\phi_i$ and $1/\phi_j$.
Given these drawbacks, Multivariate Gamma's will not be adopted here.

\subsection{Copula Based Formulations}
\label{sect:copulas}

A different approach to specify the distribution of $\bm{\varepsilon}_t|\mathcal{F}_{t-1}$ is to use copula functions (cf., among others, \citet{Joe:1997} and \citet{Nelsen:1999}, \citet{Embrechts:McNeil:Straumann:2002}, \citet{Cherubini:Luciano:Vecchiato:2004}, \citet{McNeil:Frey:Embrechts:2005} and the review of \citet{Patton:2007} for financial applications).
In this case, the conditional pdf of the error component of the vMEM is given by
\begin{equation}
  \label{eqn:pdfEpsCopula}
	f_{\bm{\varepsilon}}(\bm{\varepsilon}_t | \mathcal{F}_{t-1}) =
		c(\bm{u}_t; \bm{\xi}) \prod_{i = 1}^K f_i(\varepsilon_{t,i}; \bm{\phi}_i),
\end{equation}
where $c(\bm{u}_t; \bm{\xi})$ is the pdf of the copula, $f_i(\varepsilon_{t,i}; \bm{\phi}_i)$ and $u_{t,i} = F_i(\varepsilon_{t,i}; \bm{\phi}_i)$ are the pdf and the cdf, respectively, of the marginals, $\bm{\xi}$ and $\bm{\phi}_{i}$ are parameters.

A copula approach, hence, requires the specification of two elements: the distribution of the marginals and the copula function.
In view of the flexible properties shown elsewhere \citep{Engle:Gallo:2006}, for the first we adopt Gamma pdf's (but other choices are possible, such as Inverse-Gamma, Weibull, Lognormal, and mixtures of them).
For the second, we discuss some possible specifications within the class of Elliptical copulas.

\subsubsection{Normal Copula}
\label{sect:NormalCopula}

The Normal copula is a frequent choice in applications (\citet{McNeil:Frey:Embrechts:2005}, \citet{Cherubini:Luciano:Vecchiato:2004}, \citet{Bouye:Durrleman:Nikeghbali:Riboulet:Roncalli:2000}).
Its pdf is given by
\begin{equation}
  \label{eqn:NormalCopula}
  c_{N}(\bm{u}; \bm{R}) = |\bm{R}|^{-1/2} \exp
    \left[
    -\frac{1}{2} \left( \bm{q}^\prime \bm{R}^{-1} \bm{q} - \bm{q}^\prime \bm{q} \right)
    \right],
\end{equation}
where $\bm{q} = (q_1; \ldots; q_K)$, $q_i = \Phi^{-1}(u_i)$ and $\Phi(x)$ denotes the cdf of the standard Normal distribution computed at $x$.

The Normal copula has many interesting properties:
the ability to reproduce a broad range of dependencies (the bivariate version, according to the value of the correlation parameter, is capable of attaining the lower Fr\'echet bound, the product copula and the upper Fr\'echet bound.), the analytical tractability, the ease of simulation.
When combined with $Gamma(\phi_i, \phi_i)$ marginals, the resulting multivariate distribution is a special case of dispersion distribution generated from a Gaussian copula \citep{Song:2000}.
We note that the conditional correlation matrix of $\bm{\varepsilon}_t$ has generic element approximately equal to $R_{ij}$, which, differently from the Multivariate Gamma in Section~\ref{sect:MultivariateGamma}, can assume negative values too.

\subsubsection{Student-T Copula}
\label{sect:TCopula}

One of the limitations of the Normal copula is the asymptotic independence of its tails. Empirically, tail dependence is a behavior observed frequently in financial time series (see \citet{McNeil:Frey:Embrechts:2005}, among others). Elements of $\bm{x}$ (be they different indicators of the same asset or different assets) tend to be affected by the same extreme events.
For this reason, as an alternative, we consider the Student-T copula which allows for asymptotically dependent tails.
The pdf of the Student-T copula is given by
\begin{equation}
  \label{eqn:TCopula}
  c_{T}(\bm{u}; \bm{R}, \nu) =
    \frac{\Gamma((\nu+K)/2) \Gamma(\nu/2)^{K-1}}{\Gamma((\nu+1)/2)}
    |\bm{R}|^{-1/2}
    \frac{(1 + \bm{q}^\prime \bm{R}^{-1} \bm{q} / \nu)^{-(\nu + K)/2}}
    {\prod_{i = 1}^K (1 + q_i^2 / \nu)^{-(\nu + 1)/2}},
\end{equation}
where $\bm{q} = (q_1; \ldots; q_K)$, $q_i = T^{-1}(u_i; \nu)$ and $T(x; \nu)$ denotes
the cdf of the Student-T distribution with $\nu$ degrees of freedom computed at $x$.
Differently from the Normal copula, when $\bm{R} = \bm{I}$ we get uncorrelated but not independent marginals (details in \citet{McNeil:Frey:Embrechts:2005}).
Further specifications of the Student-T copula are in \citet{Demarta:McNeil:2005}.

\subsubsection{Elliptical Copulas}
\label{sect:EllipticalCopula}

The Elliptical copulas family provides an unified framework encompassing the Normal, the Student-T and any other member of this family endowed of an explicit pdf.
Elliptical copulas (for details see \citet{McNeil:Frey:Embrechts:2005}, \citet{Frahm:Junker:Szimayer:2003},
\citet{Schmidt:2002}) are copulas generated by Elliptical distributions, exactly in the same way as the Normal copula and the Student-T copula
stem from the multivariate Normal and Student-T distributions, respectively. Elliptical copulas have interesting features and widespread applicability, even if their elliptical symmetry may constitute a limit in some applications.\footnote{Copulas in the Archimedean family offer a way to bypass such a limitation but suffer from other drawbacks and will not be pursued here.}

We consider a copula generated by an Elliptical distribution whose univariate 'standardized' marginals (intended here with location parameter $0$ and dispersion parameter $1$) have an absolutely continuous symmetric distribution, centered at zero, with pdf $g(.; \bm{\nu})$ and cdf $G(.; \bm{\nu})$ ($\bm{\nu}$ represents a vector of shape parameters).
The density of the copula can then be written as
\begin{equation}
  \label{eqn:pdfEpsEllCopula}
  c_{E}(\bm{u}; \bm{R}, \bm{\nu}) =
  \displaystyle
  K^*(\bm{\nu}, K)
  |R|^{-1/2}
  \frac{g_1(\bm{q}^\prime \bm{R}^{-1} \bm{q}; \bm{\nu}, K)}
  {\prod_{i = 1}^K g_2(q_i^2; \bm{\nu})}
\end{equation}
for suitable choices of $K^*(.,.)$, $g_1(.;.,.)$ and $g_2(.;.)$, where
$\bm{q} = (q_1; \ldots; q_K)$,
$q_i = G^{-1}(u_i; \bm{\nu})$.
For instance:
\begin{itemize}
	\item in the Normal copula, with no explicit shape parameter $\bm{\nu}$,
				we have $K^*(K) \equiv 1$, $g_1(x; K) = g_2(x) \equiv \exp(-x / 2)$;
	\item in the Student-T copula, with a scalar $\nu$ shape parameter, we have
				$K^*(\nu; K) = \displaystyle \frac{\Gamma((\nu + K)/2) \Gamma(\nu/2)^{K-1}}{\Gamma((\nu + 1)/2)}$,
				$g_1(x; \nu, K) = (1 + x / \nu)^{-(\nu + K)/2}$, $g_2(x; \nu) = (1 + x / \nu)^{-(\nu + 1)/2}$.
\end{itemize}

\section{Maximum Likelihood Inference}
\label{sect:MLInference}

In this section we discuss how to get full Maximum Likelihood (ML) inferences from the vMEM with the parametric specification (\ref{eqn:muVectorL}) for $\bm{\mu}_t$ (dependent on a parameter vector $\bm{\theta}$) and a generic formulation $f_{\bm{\varepsilon}}(\bm{\varepsilon}_t | \mathcal{F}_{t-1})$ of the conditional distribution of the vector error term (characterized by the parameter vector $\bm{\lambda}$).
Inference on $\bm{\theta}$ and $\bm{\lambda}$ can be discussed in turn, given that from the model assumptions the log-likelihood function is
\begin{eqnarray}
  \label{eqn:logLik}
  l
  &=&
  \sum_{t = 1}^T \ln f_{\bm{x}}(\bm{x}_t | \mathcal{F}_{t-1}) =
  \sum_{t=1}^T \ln \left(f_{\bm{\varepsilon}}(\bm{\varepsilon}_t | \mathcal{F}_{t-1})
  \prod_{i = 1}^K \mu_{t,i}^{-1} \right) \nonumber{}
  \\
  &=&
  \sum_{t = 1}^T \left[ \ln f(\bm{\varepsilon}_t | \mathcal{F}_{t-1}) - \sum_{i = 1}^K \ln \mu_{t,i} \right].
\end{eqnarray}

Considering a generic time $t$, it is useful to recall the sequence of calculations:
\begin{equation*}
	\mu_{t,i}(\bm{\theta}_i)
	\rightarrow x_{t,i} / \mu_{t,i} = \varepsilon_{t,i} \rightarrow
	F_i(\varepsilon_{t,i}; \bm{\phi}_i) = u_{t,i} \rightarrow
	c(\bm{u}_t; \bm{\xi})
	\qquad
	i = 1, \ldots, K
\end{equation*}
where $\bm{\theta}_{i}$ is the parameter involved in the $i$-th element of the $\bm{\mu}_{t}$ vector.

\subsection{Parameters in the Conditional Mean}
\label{sect:MLInferenceMu}

Irrespective of the specification chosen for $f_{\bm{\varepsilon}}(\bm{\varepsilon}_t | \mathcal{F}_{t-1})$, the structure of the vMEM allows to express  the portion of the score function corresponding to $\bm{\theta}$ as
\begin{equation}
  \label{eqn:scoreTheta}
  \nabla_{\bm{\theta}}\, l
    =
  \sum_{t = 1}^T \bm{A}_t \bm{w}_t
\end{equation}
where
\begin{equation*}
  \bm{A}_t = - \nabla_{\bm{\theta}} \bm{\mu}_t^\prime \diag(\bm{\mu}_t)^{-1}.
\end{equation*}
\begin{equation} \label{eqn:w}
  \bm{w}_t = \bm{\varepsilon}_t \odot \bm{b}_t + \mathds{1},
\end{equation}
\begin{equation*}
  \bm{b}_t = \nabla_{\bm{\varepsilon}_t} \ln f(\bm{\varepsilon}_t | \mathcal{F}_{t-1}),
\end{equation*}
In order to have a zero expected score, we need $E(\bm{w}_t | \mathcal{F}_{t-1}) = \bm{0}$ or, equivalently,
$E(\bm{\varepsilon}_t \odot \bm{b}_t | \mathcal{F}_{t-1}) = -\mathds{1}$.
As a consequence, the information matrix and the expected Hessian are given by
\begin{equation}
  E \left[ \bm{A}_t \bm{\mathcal{I}}^{(\varepsilon)} \bm{A}_t^\prime \right]
\end{equation}
and
\begin{equation}
  E \left[ \bm{A}_t \bm{H}^{(\varepsilon)} \bm{A}_t^\prime \right],
\end{equation}
respectively, where the matrices
\begin{equation*}
  \bm{\mathcal{I}}^{(\varepsilon)}
    =
  E \left[(\bm{\varepsilon}_t \odot \bm{b}_t) (\bm{\varepsilon}_t \odot \bm{b}_t)^\prime
    | \mathcal{F}_{t-1} \right]
    -
  \mathds{1} \mathds{1}^\prime
\end{equation*}
and
\begin{equation*}
  \bm{H}^{(\varepsilon)}
    =
  E \left[
    \nabla_{\bm{\varepsilon}_t} \bm{b}_t^\prime (\bm{\varepsilon}_t \bm{\varepsilon}_t^\prime)
    | \mathcal{F}_{t-1}
  \right]
    -
  \bm{I}
\end{equation*}
depend only on $\bm{\lambda}$ but not on $\bm{\theta}$. Of course, under a correct specification $\bm{H}^{(\varepsilon)} = - \bm{\mathcal{I}}^{(\varepsilon)}$.

For a particular parametric choice of the conditional distribution of $\bm{\varepsilon}_t$, we need to plug the specific expression of
$\ln f_{\bm{\varepsilon}}(\bm{\varepsilon}_t | \mathcal{F}_{t-1})$ into $\bm{b}_{t}$.
For instance, considering the generic copula formulation (\ref{eqn:pdfEpsCopula}), then
\begin{equation}\label{eqn:lnpdfEpsCopula}
  \ln f_{\bm{\varepsilon}}(\bm{\varepsilon}_t | \mathcal{F}_{t-1})
    =
  \ln c(\bm{u}_t) + \sum_{i = 1}^{K} \ln f_i(\varepsilon_{t,i})
\end{equation}
so that $\bm{b}_t$ has elements
\begin{equation}
  b_{t,i}
    =
  f_{i}(\varepsilon_{t,i}) \nabla_{u_{t,i}} \ln c(\bm{u}_t)
    +
  \nabla_{\varepsilon_{t,i}} \ln f_i(\varepsilon_{t,i}).
\end{equation}
In what follows we provide specific formulas for the elliptical copula formulation (\ref{eqn:pdfEpsEllCopula}) and its main sub-cases.

\subsection{Parameters in the pdf of the Error Term}
\label{sect:MLInferenceEps}

Under a copula approach, the portion of the score function corresponding to the term $\sum_{t = 1}^T  \ln f(\bm{\varepsilon}_t | \mathcal{F}_{t-1})$ (cf.~Equation~(\ref{eqn:logLik})) depends on a vector $\bm{\lambda} = (\bm{\xi}; \bm{\phi})$ ($\bm{\xi}$ and $\bm{\phi}$ are the parameters of the copula function and of the marginals, respectively -- cf.~Section~\ref{sect:copulas}),
\begin{equation*}
  \nabla_{\bm{\lambda}}\, l
    =
  \sum_{t = 1}^T \nabla_{\bm{\lambda}} \ln f(\bm{\varepsilon}_t | \mathcal{F}_{t-1}) = \sum_{t = 1}^T \nabla_{\bm{\lambda}} \ln \left(c(\bm{u}_t; \bm{\xi}) \prod_{i = 1}^K f_i(\varepsilon_{t,i}; \bm{\phi}_i)\right).
\end{equation*}
Therefore,
\begin{equation*}
  \nabla_{\bm{\xi}}\, l
    =
  \sum_{t = 1}^T \nabla_{\bm{\xi}} \ln c(\bm{u}_t).
\end{equation*}
and
\begin{equation*}
  \nabla_{\bm{\phi}_i}\, l
    =
  \sum_{t = 1}^T
    \left[ \nabla_{\bm{\phi}_i} F_i(\varepsilon_{t,i}) \nabla_{\bm{u}_{t,i}} \ln c(\bm{u}_t)
      +
    \nabla_{\bm{\phi}_i} \ln f_i(\varepsilon_{t,i}) \right].
\end{equation*}
As detailed in Section \ref{sect:EllipticalCopula}, beside a possible shape parameter $\bm{\nu}$, elliptical copulas are characterized
by a correlation matrix $\bm{R}$ which, in view of its full ML estimation, can be expressed (cf. \citet[p.~235]{McNeil:Frey:Embrechts:2005}) as
\begin{equation} \label{eqn:RRepresentation}
   \bm{R} = \bm{D}\bm{c}^\prime \bm{c} \bm{D},
\end{equation}
where $\bm{c}$ is an upper-triangular matrix with ones on the main diagonal and $\bm{D}$ is a diagonal matrix with diagonal entries
$D_1 = 1$ and $D_j = \left(1 + \sum_{i = 1}^{j-1} c_{ij}^2 \right)^{-1/2}$ for $j = 2, \ldots, K$.
So doing, the estimation of $\bm{R}$ is transformed in an unconstrained problem, since the $K(K-1)/2$ free elements of $\bm{c}$ can vary into $\R$.
We can then write $\bm{\xi} = (\bm{c}; \bm{\nu})$.

Let us introduce a compact notation as follows:
$\bm{C} = \bm{c D}$, $\bm{q}_t = (q_{t,1}; \ldots; q_{t,K})$, $q_{t,i} = G^{-1}(u_{t,i}; \bm{\nu})$, $\widetilde{\bm{q}}_t = \bm{C}^{\prime -1} \bm{q}_t$, $\bm{q}^*_t = \bm{R}^{-1} \bm{q}_t$, $\widetilde{\widetilde{q}}_t
= \bm{q}_t^\prime \bm{R}^{-1} \bm{q}_t =
\widetilde{\bm{q}}_t^\prime \widetilde{\bm{q}}_t$.
We can then write
\begin{equation} \label{eqn:EllipticalCLL}
   \ln c(\bm{u}_t) =
  \ln K^* - \sum_{i = 2}^K \ln{D_i}
    +
  \ln g_1(\widetilde{\widetilde{q}}_t)
    -
  \sum_{i = 1}^K \ln g_2 (q_{t,i}^2)
\end{equation}
where use was made of the fact that $\frac{1}{2} \ln(|\bm{R}|) =
\sum_{i = 2}^K \ln{D_i}$.

In specific cases we get:
\begin{itemize}
	\item Normal copula: $\ln K^* = 0 $, $\ln g_1(x) = \ln g_2(x) = -x / 2$;
	\item Student-T copula:
				$\ln K^* = \ln \left[ \frac{\Gamma((\nu + K)/2) \Gamma(\nu/2)^{K-1}}{\Gamma((\nu + 1)/2)} \right]$,
				$\ln g_1(x) = -\frac{\nu + K}{2} \ln \left(1 + \frac{x}{\nu} \right)$,
				$g_2(x) = -\frac{\nu + 1}{2} \ln \left(1 + \frac{x}{\nu} \right)$.
\end{itemize}

\vspace{1em}
\textbf{Parameters entering the matrix $\bm{c}$}

The portion of the score relative to the free parameters of the $\bm{c}$ matrix has elements
\begin{equation} \label{eqn:derlderc}
  \nabla_{c_{ij}}\, l
    =
  \nabla_{c_{ij}} \left[ - T \sum_{i = 2}^K \ln D_i
    + \sum_{t = 1}^T \ln g_1(\widetilde{\widetilde{q}}_t)\right], \ i<j.
\end{equation}
Using some algebra we can show that
\begin{equation*}
   \Der{c_{ij}}{\sum_{i = 2}^K \ln (D_i)} = - D_j C_{ij} \\
\end{equation*}
and
\begin{equation*}
   \nabla_{c_{ij}}\ln g_1(\widetilde{\widetilde{q}}_t) =
     -2 \Der{\widetilde{\widetilde{q}}_t}{(\ln g_1(\widetilde{\widetilde{q}}_t)) D_j q_{t,j}^* (\widetilde{q}_{t,i} - C_{ij} q_{t,j})}.
\end{equation*}
By replacing them into (\ref{eqn:derlderc}) we obtain
\begin{equation*}
   \Der{c_{ij}}{l} = T D_j C_{ij} +
                     2 D_j \sum_{t = 1}^T q_{t,j}^*
                     (C_{ij} q_{t,j} - \widetilde{q}_{t,i})
                     \Der{\widetilde{\widetilde{q}}_t}{(\ln g_1(\widetilde{\widetilde{q}}_t))}.
\end{equation*}

\vspace{1em}
\textbf{Parameters entering the vector $\bm{\nu}$}

The portion of the score relative to $\bm{\nu}$ is
\begin{align*}
   \nabla_{\bm{\nu}}l =  \Der{\bm{\nu}}{\left[T \ln K^*
                      + \sum_{t = 1}^T \ln g_1(\widetilde{\widetilde{q}}_t)
                      - \sum_{t = 1}^T \sum_{i = 1}^K \ln g_2 (q_{t,i}^2) \right]}.
\end{align*}
The derivative of $\ln K^* = \ln K^*(\bm{\nu}; K)$ can sometimes be computed
analytically. For instance, in the Student--T copula we have
\begin{align*}
   \Der{\nu}{\ln K^*(\nu; K)} = \frac{1}{2} \left[\psi\left(\frac{\nu + K}{2}\right)
                                                       + (K-1)\psi\left(\frac{\nu}{2}\right)
                                                       - K \psi\left(\frac{\nu + 1}{2}\right) \right].
\end{align*}
For the remaining quantities we suggest numerical derivatives when, as in the Student--T case, the quantile function $G^{-1}(x;\bm{\nu})$ cannot be computed analytically.

\vspace{1em}
\textbf{Parameters entering the vector $\bm{\phi}$}

The portion of the score relative to $\bm{\phi}$ has elements
\begin{align*}
 \Der{\bm{\phi_i}}{l} = \Der{\bm{\phi_i}}{}
			 \sum_{t = 1}^T \left[
												\ln g_1(\widetilde{\widetilde{q}}_t)
												- \sum_{i = 1}^K \ln g_2 (q_{t,i}^2)
												+ \sum_{i = 1}^K \ln f_i (\varepsilon_{t,i})
											\right].
\end{align*}
After some algebra we obtain
\begin{equation} \label{eqn:DerPhiEC}
  \Der{\bm{\phi_i}}{l}
    =
  \sum_{t = 1}^T
	\left[
		\nabla_{\bm{\phi}_i} F_i(\varepsilon_{t,i}) d_{t,i} + \nabla_{\bm{\phi}_i} \ln f_i(\varepsilon_{t,i})
	\right],
\end{equation}
where
\begin{equation} \label{eqn:dti}
  d_{t,i}
    =
  \frac{1}{g(q_{t,i})}
    \left[
  2 q_{t,i}^* \nabla_{\widetilde{\widetilde{q}}_{t}} \ln g_1(\widetilde{\widetilde{q}}_{t})
    -
  \nabla_{q_{t,i}} \ln g_2(q_{t,i}^2)
  \right].
\end{equation}
For instance, if a
marginal has a distribution $Gamma(\phi_i, \phi_i)$ then
\begin{align*}
   \Der{\phi_i}{f_i(\varepsilon_{t,i})} = \ln(\phi_i) - \psi(\phi_i) + \ln(\varepsilon_{t,i}) - \varepsilon_{t,i} + 1,
\end{align*}
whereas $\Der{\phi_i}{F_i(\varepsilon_{t,i})}$ can be computed
numerically.

\vspace{1em}
\textbf{Parameters entering the vector $\bm{\theta}$}

By exploiting the notation introduced in this section, we can now detail the structure of $\bm{b}_t$ entering
into (\ref{eqn:w}) and then into the portion of the score function relative to $\bm{\theta}$.
From (\ref{eqn:EllipticalCLL}), $\bm{\varepsilon}_t \odot \bm{b}_t + \mathds{1}$ (cf.~\ref{eqn:w}) has elements
\begin{equation}
  \varepsilon_{t,i} b_{t,i} +1
    =
  \varepsilon_{t,i} f_{i}(\varepsilon_{t,i}) d_{t,i}
    +
  \varepsilon_{t,i} \nabla_{\varepsilon_{t,i}} \ln f_i(\varepsilon_{t,i})) + 1
\end{equation}
where $d_{t,i}$ is given in (\ref{eqn:dti}).
For our choice, $f_i(\varepsilon_{t,i})$ is the pdf of a $Gamma(\phi_i, \phi_i)$ distribution, so that
\begin{equation} \label{eqn:dGamma:dPhi}
   \varepsilon_{t,i}\nabla_{\varepsilon_{t,i}} \ln f_i(\varepsilon_{t,i}) +1 = \phi_i - \varepsilon_{t,i} \phi_i .
\end{equation}

\subsubsection{Expectation Targeting}
\label{sect:ExpectationTargeting}

Assuming weak-stationarity of the process, numerical stability and a reduction in the number of parameters to be estimated can be achieved by expressing  $\bm{\omega}$ in terms of the unconditional mean of the process, say $\bm{\mu}$, which can be easily estimated by the sample mean (\textit{expectation targeting}%
\footnote{%
This is equivalent to \textit{variance targeting} in a GARCH context (\citet{Engle:Mezrich:1995}), where the constant term  of the conditional variance model is assumed to be a function of the sample unconditional variance and of the other parameters.
In this context, other than a preference for the term expectation targeting since we are modeling a conditional mean, the main argument stays unchanged.}).
Since $E(\bm{x}_t) = E(\bm{\mu}_t) = \bm{\mu}$ is well defined and is equal to
\begin{equation}
  \label{eqn:mu}
  \bm{\mu}
    =
	\left[
	  \bm{I} - \sum_{l=1}^{L} \left( \bm{\alpha}_l + \bm{\beta}_l + \frac{\bm{\gamma}_l}{2} \right)
	\right]^{-1}
	\bm{\omega},
\end{equation}
Equation~(\ref{eqn:muVectorL}) becomes
\begin{equation}
  \label{eqn:muVectorL:ET}
  \bm{\mu}_t
    =
  \left[
	  \bm{I} - \sum_{l=1}^{L} \left( \bm{\alpha}_l + \bm{\beta}_l + \frac{\bm{\gamma}_l}{2} \right)
	\right] \bm{\mu}
	  +
	\sum_{l=1}^{L} \left( \bm{\alpha}_l \bm{x}_{t-l} + \bm{\gamma}_l \bm{x}_{t-l}^{(-)} +  \bm{\beta}_l \bm{\mu}_{t-l}\right).
\end{equation}
In a GARCH framework, the consequences of this practice have been been investigated by~\citet{Kristensen:Linton:2004} and, more recently, by \citet{Francq:Horvath:Zakoian:2011} who show its merits in terms of stability of estimation algorithms and accuracy in estimation of both coefficients and long term volatility (cf. Appendix \ref{app:exptarg} for some details in the present context).

From a technical point of view, a trivial replacement of $\bm{\mu}$ by the sample mean $\overline{\bm{x}}_T$ followed by a ML estimation of the remaining parameters, preserves consistency but leads to wrong standard errors \citep{Kristensen:Linton:2004}.
The issue can faced by identifying the inference problem as involving a two-step estimator (\citet[ch.~6]{Newey:McFadden:1994}), namely by rearranging $(\bm{\theta};\bm{\lambda})$ as $(\bm{\mu}; \bm{\vartheta})$, where $\bm{\vartheta}$ collects all model parameters but $\bm{\mu}$.
Under conditions able to guarantee consistency and asymptotic normality of $(\bm{\vartheta}; \bm{\mu})$ (in particular, the existence of $E(\bm{\mu}_t \bm{\mu}_t^\prime)$: see \citet{Francq:Horvath:Zakoian:2011}), we can adapt the notation of \citet{Newey:McFadden:1994} to write the asymptotic variance of $\sqrt{T} \left(\widehat{\bm{\vartheta}}_T - \bm{\vartheta}\right)$ as
\begin{equation}
  \label{eqn:AsyVar:vartheta}
  \bm{G}_{\bm{\vartheta}}^{-1}
	\left[
	\begin{array}{cc}
		\bm{I}  &  -\bm{G}_{\bm{\mu}} \bm{M}^{-1}
	\end{array}
	\right]
  \left[
	\begin{array}{cc}
		\bm{\Omega}_{\bm{\vartheta}, \bm{\vartheta}^\prime}  &  \bm{\Omega}_{\bm{\vartheta},\bm{\mu}^\prime} \\
		\bm{\Omega}_{\bm{\mu},\bm{\vartheta}^\prime}         &  \bm{\Omega}_{\bm{\mu},\bm{\mu}^\prime} \\
	\end{array}
	\right]
	\left[ \begin{array}{c}
		\bm{I}
			\\
		-\left(\bm{G}_{\bm{\mu}} \bm{M}^{-1} \right)^\prime
	\end{array} \right]
  \bm{G}_{\bm{\vartheta}}^{-1 \prime}
\end{equation}
where
\begin{align*}
  \bm{G}_{\bm{\vartheta}} & = E(\nabla_{\bm{\vartheta} \bm{\vartheta}^\prime} l_t) \\
  \bm{G}_{\bm{\mu}}       & = E(\nabla_{\bm{\vartheta} \bm{\mu}^\prime} l_t)       \\
  \bm{M}                  & = E(\nabla_{\bm{\mu}^\prime} \bm{m}_t),
\end{align*}
and
\begin{equation*}
  \bm{m} = \sum_{t = 1}^T \bm{m}_t = \sum_{t = 1}^T \left( \bm{x}_t - \bm{\mu} \right)
\end{equation*}
is the moment function giving the sample average $\overline{\bm{x}}_T$ as an estimator of $\bm{\mu}$. The $\bm{\Omega}$ matrix denotes the variance-matrix of $\left( \nabla_{\bm{\vartheta}} l_t; \bm{m}_t \right)$ partitioned in the corresponding blocks.

To give account of the necessary modifications to adopt with expectation targeting, we provide sufficiently general and compact expressions for the parameters $\bm{\mu}$ (the unconditional mean) and $\bm{\theta}$ (the remaining parameters) in the conditional mean expressed by~(\ref{eqn:muVectorL:ET}).
\begin{align*}
  \bm{G}_{\bm{\theta}} & = E \left( \nabla_{\bm{\theta} \bm{\theta}^\prime} l_t \right) = E \left[ \bm{A}_t \bm{H}^{(\varepsilon)} \bm{A}_t^\prime \right]
	  \\
  \bm{G}_{\bm{\mu}} & = E(\nabla_{\bm{\theta} \bm{\mu}^\prime} l_t) = - E \left( \bm{A}_t \bm{H}^{(\varepsilon)} \diag(\bm{\mu_t})^{-1} \right) \bm{A}
    \\
	\bm{M} & = -\bm{I}
	  \\
  \bm{\Omega}_{\bm{\theta}, \bm{\theta}^\prime}  & = E \left( \nabla_{\bm{\theta}} l_t \nabla_{\bm{\theta}^\prime} l_t \right)
    =
  E \left[ \bm{A}_t \bm{\mathcal{I}}^{(\varepsilon)} \bm{A}_t^\prime \right]
    \\
  \bm{\Omega}_{\bm{\theta}, \bm{\mu}^\prime} &
    =
  E \left( \nabla_{\bm{\theta}} l_t \bm{m}_t^\prime \right)
    =
  E \left[ \bm{A}_t \left[ E \left[ \left( \bm{b}_t \odot \bm{\varepsilon}_t \right) \bm{\varepsilon}_t^\prime | \mathcal{F}_{t-1} \right]  + \mathds{1} \mathds{1}^\prime\right] \diag(\bm{\mu}_t) \right] \bm{B} \bm{A}^{-1 \prime}
    \\
	\bm{\Omega}_{\bm{\mu},\bm{\mu}^\prime}
	  & =
	E \left( \bm{m}_t \bm{m}_t^\prime \right)
	  =
	\bm{A}^{-1}
	 \left( \bm{B} \bm{\Sigma}_v \bm{B}^\prime  + \bm{C} \left( \bm{\Sigma}_v \odot \bm{\Sigma}_I \right) \bm{C}^\prime  \right)
	\bm{A}^{-1 \prime}
\end{align*}
where
\begin{align*}
	\bm{A} & = \bm{I} - \sum_{l=1}^{L} \left( \bm{\alpha}_l + \bm{\beta}_l + \frac{\bm{\gamma}_l}{2} \right) \\
	\bm{B} & = \bm{I} - \sum_{l=1}^{L} \bm{\beta}_l \\
	\bm{C} & = \sum_{l=1}^{L} \bm{\gamma}_l \\
	\bm{\Sigma}_v & = E(\bm{\mu}_t \bm{\mu}_t^\prime) \odot \bm{\Sigma}.
\end{align*}
The expression for $\bm{\Omega}_{\bm{\mu},\bm{\mu}^\prime}$ is obtained by using the technique in \citet{Horvath:Kokozka:Zitikis:2006} and   \citet{Francq:Horvath:Zakoian:2011}.
In this sense, we extend the cited works to a multivariate formulation including asymmetric effects as well.

Some further simplification is also possible when the model is correctly specified since, in such a case, $\bm{H}^{(\varepsilon)} = -\bm{\mathcal{I}}^{(\varepsilon)}$ and
$E \left[ \left( \bm{b}_t \odot \bm{\varepsilon}_t \right) \bm{\varepsilon}_t^\prime | \mathcal{F}_{t-1} \right]  = - \mathds{1} \mathds{1}^\prime$ leading to $\bm{\Omega}_{\bm{\theta}, \bm{\mu}^\prime} = \bm{0}$ and to
\begin{equation*}
	E \left[ \bm{A}_t \bm{\mathcal{I}}^{(\varepsilon)} \bm{A}_t^\prime \right]
	  +
	E \left( \bm{A}_t \bm{\mathcal{I}}^{(\varepsilon)} \diag(\bm{\mu_t})^{-1} \right)
	\left[
	\bm{\Sigma}_v \bm{B}^\prime  + \bm{C} \left( \bm{\Sigma}_v \odot \bm{\Sigma}_I \right) \bm{C}^\prime
	\right]
	E \left( \diag(\bm{\mu_t})^{-1} \bm{\mathcal{I}}^{(\varepsilon)} \bm{A}_t^\prime \right)
\end{equation*}
for what concerns the inner part of (\ref{eqn:AsyVar:vartheta}).

\subsubsection{Concentrated Log-likelihood}
\label{sect:ConcentratedLogLik}

Some further numerical estimation stability and reduction in the number of parameters can be achieved -- if needed -- within the framework of elliptical copulas: we can use current values of residuals to compute current estimates of $\bm{R}$ (Kendall correlations are suggested by \citet{Lindskog:McNeil:Schmock:2003}) and of the shape parameter $\bm{\nu}$ (tail dependence indices are proposed by \citet{Kostadinov:2005}).
This approach may be fruitful with other copulas as well when sufficiently simple moment conditions can be exploited.
A similar strategy can be applied also to the parameters of the marginals. For instance, if they are assumed $Gamma(\phi_i, \phi_i)$ distributed, the relation $V(\varepsilon_{t,i} | \mathcal{F}_{t-1}) = 1 / \phi_i$ leads to very simple estimator of $\phi_i$ from current values of residuals.
By means of this approach, the remaining parameters can be updated from a sort of pseudo-loglikelihood conditioned on current estimates of the pre--estimated parameters.

In the case of a Normal copula a different strategy can be followed. The formula of the (unconstrained) ML estimator of the $\bm{R}$ matrix (\citet[p.~155]{Cherubini:Luciano:Vecchiato:2004}), namely
\begin{equation*}
  \bm{Q} = \frac{\bm{q}^\prime \bm{q}}{T}
\end{equation*}
where $\bm{q} = (\bm{q}_1^\prime;\ldots;\bm{q}_T^\prime)$ is a $T\times K$ matrix, can be plugged into the log-likelihood in place of $\bm{R}$ obtaining a sort of \textit{concentrated log-likelihood}
\begin{equation} \label{eqn:NormalCCLL1}
  \frac{T}{2} \left[ - \ln |\bm{Q}| - K + \trace(\bm{Q}) \right] + \sum_{t = 1}^T \sum_{i = 1}^K \ln f_i(\varepsilon_{t,i} | \mathcal{F}_{t-1})
\end{equation}
leading to a relatively simple structure of the score function.
However, this estimator of $\bm{R}$ is obtained without imposing any constraint relative to its nature as a correlation matrix ($\diag(\bm{R}) = \mathds{1}$ and positive definiteness).
Computing directly the derivatives with respect to the off--diagonal elements of $\bm{R}$ we obtain, after some algebra, that the constrained ML estimator of $\bm{R}$ satisfies the following equations:
\begin{align*}
    (\bm{R}^{-1})_{ij}-(\bm{R}^{-1})_{i.}\frac{\bm{q}^\prime \bm{q}}{T}(\bm{R}^{-1})_{.j} = 0
\end{align*}
for $i \neq j = 1,\ldots,K$, where $\bm{R}_{i.}$ and $\bm{R}_{.j}$ indicate, respectively, the $i$--th row and the $j$--th
column of the matrix $\bm{R}$.
Unfortunately, these equations do not have an explicit solution.\footnote{Even when $\bm{R}$ is a $(2,2)$ matrix, the value of $\bm{R}_{12}$ has to satisfy the cubic equation:
\begin{align*}
    \bm{R}_{12}^3
    - \bm{R}_{12}^2\frac{q_1^\prime q_2}{T}
    + \bm{R}_{12}\left[\frac{q_1^\prime q_1}{T}
    + \frac{q_2^\prime q_2}{T}-1\right]
    - \frac{q_1^\prime q_2}{T}
    = 0.
\end{align*}
}
An acceptable compromise which should increase efficiency, although formally it cannot be interpreted as an ML estimator, is to normalize the estimator $\bm{Q}$ obtained above in order to transform it to a correlation matrix:
\begin{equation*}
    \widetilde{\bm{R}}  = \bm{D}_Q^{-\frac{1}{2}} \bm{Q} \bm{D}_Q^{-\frac{1}{2}},
\end{equation*}
where $\displaystyle \bm{D}_Q = \diag( Q_{11}, \ldots, Q_{KK})$.
This solution can be justified observing that the copula contribution to the likelihood depends on $\bm{R}$ exactly as if it was the correlation matrix of iid~rv's~$\bm{q}_t$ normally distributed with mean $\bm{0}$ and correlation matrix $\bm{R}$ (see also \citet[p.~235]{McNeil:Frey:Embrechts:2005}).
Using this constrained estimator of $\bm{R}$, the concentrated log-likelihood becomes
\begin{equation} \label{eqn:NormalCCLL2}
   \frac{T}{2} \left[ - \ln |\widetilde{\bm{R}}|
           - \trace( \widetilde{\bm{R}}^{-1} \bm{Q}
           + \trace(\bm{Q}) \right]
           + \sum_{t = 1}^T \sum_{i = 1}^K \ln f_i(\varepsilon_{t,i} | \mathcal{F}_{t-1}).
\end{equation}
It is interesting to note that, as long as (\ref{eqn:NormalCCLL1}), Equation~(\ref{eqn:NormalCCLL2}) too gives a relatively simple structure of the score
function.
Using some tedious algebra, we can show that the components of the score corresponding to $\bm{\theta}$ and $\bm{\phi}$ have exactly the same structure as above, with the quantity $d_{t,i}$ into (\ref{eqn:dti}) changed to
\begin{align} \label{eqn:dti2}
   d_{t,i} = \frac{\bm{C}_{i.} \bm{q}_t}{\phi(q_{t,i})}
\end{align}
where the $\bm{C}$ matrix is here given by
\begin{align*}
	\bm{C} = \bm{Q}^{-1} \bm{D}_Q^{1/2} \bm{Q} \bm{D}_Q^{1/2} \bm{Q}^{-1}
					- \bm{Q}^{-1} + \bm{I}_K
					- \widetilde{\bm{R}}^{-1} + \bm{D}_Q^{-1}
					- \bm{D}_Q^{-1/2} \diag(\bm{Q}^{-1} \bm{D}_Q^{1/2} \bm{Q})
\end{align*}
and $\phi(.)$ indicates here the pdf of the standard normal computed at its argument.
Of course, also in this case the parameters of the marginals can be updated by means of moment estimators computed from current residuals (instead that via ML) exactly as explained above.

\section{Trading Activity and Volatility within a vMEM} \label{sect:TradingActivity}

Trading activity produces a lot of indicators which are therefore characterized by interdependence, even in a dynamic sense. In this application, we concentrate on the joint dynamics of three series proxying such trading activity, namely volatility (measured as realized kernel volatility, cf. \citet{BarndorffNielsen:Hansen:Lunde:Shephard:2011}, and references therein), volume of shares traded and number of shares per day.
The relationship between volatility and volume as relating to trading activity was documented, for example, in the early contribution by \citet{Andersen:1996}.
Ever since, the evolution of the structure of financial markets, industry innovation, the increasing participation of institutional investors and the adoption of automated trading practices have strengthened such a relationship, and the number of trades clearly reflects an important aspect of trading intensity. To be clear at the outset, for the sake of parsimony, we choose the realized volatility forecasts as the main object of interest of the multivariate effort; we take univariate modeling of volatility by itself as a benchmark against specifications which explore the extra information in the other indicators as well as the contemporaneous correlation in the error terms.

The availability of ultra high frequency data allows us to construct daily series of the variables exploiting the most recent development in the volatility  measurement literature.
As a leading example, we consider Johnson \& Johnson (JNJ) between January 3, 2007 to July 31, 2013 ($1656$ observations). Such a stock has desirable properties of liquidity and a limited riskiness represented by a market beta generally smaller than $1$.
Raw trade data from TAQ are cleaned according to the~\citet{Brownlees:Gallo:2006} algorithm.
Subsequently, we build the daily series of realized kernel volatility, following~\citet{BarndorffNielsen:Hansen:Lunde:Shephard:2011}, computing the value at day $t$ as
\begin{equation*}
  rkv = \sqrt{ \sum_{h = -H}^{H} k \left( \frac{h}{H} \right) \gamma_{h} }
  \qquad
  \gamma_{h} = \sum_{j = |h|+1}^{n} x_{j} x_{j - |h|}
\end{equation*}
where $k(x)$ is the Parzen kernel
\begin{equation*}
  k(x) =
  \left\{
  \begin{array}{ll}
    1 - 6 x^{2} + 6 x^{3} & \text{if } x \in [0, 1/2] \\
    2 (1 - x)^{3}         & \text{if } x \in (1/2, 1] \\
    0                     & \text{otherwise}
  \end{array}
  \right.,
\end{equation*}
\begin{equation*}
  H = 3.51 \cdot n^{3/5} \left( \frac{\sum_{j = 1}^{n} x_{j}^{2} / (2n)}{ \sum_{j = 1}^{\widetilde{n}} \widetilde{x}_{j}^{2} } \right)^{2/5},
\end{equation*}
$x_{j}$ is the $j$-th high frequency return computed according to \citet[Section~2.2]{BarndorffNielsen:Hansen:Lunde:Shephard:2009}) and $\widetilde{x}_{j}$ is the intradaily return of the $j$-th bin (equally spaced on $15$ minute intervals).
For volumes ($vol$) and the number of trades ($nt$) we simply aggregate the data (sum of intradaily volumes and count of the number of trades, respectively).

\begin{figure}
  \centering
  \caption{Time series of the trading activity indices for JNJ (Jan.~3, 2007 -- July 31, 2013).
  Left: original data; Right: detrended data;
  Top: realized kernel volatility (annualized percentage); Middle: volumes (millions); Bottom: number of trades (thousands). The spike on June 13, 2012 corresponds to an important acquisition and a buy back of some of its common stock by a subsidiary.}
  \label{fig:JNJ-series}
  \begin{subfigure}[b]{0.49\textwidth}
    \includegraphics[width=\textwidth]{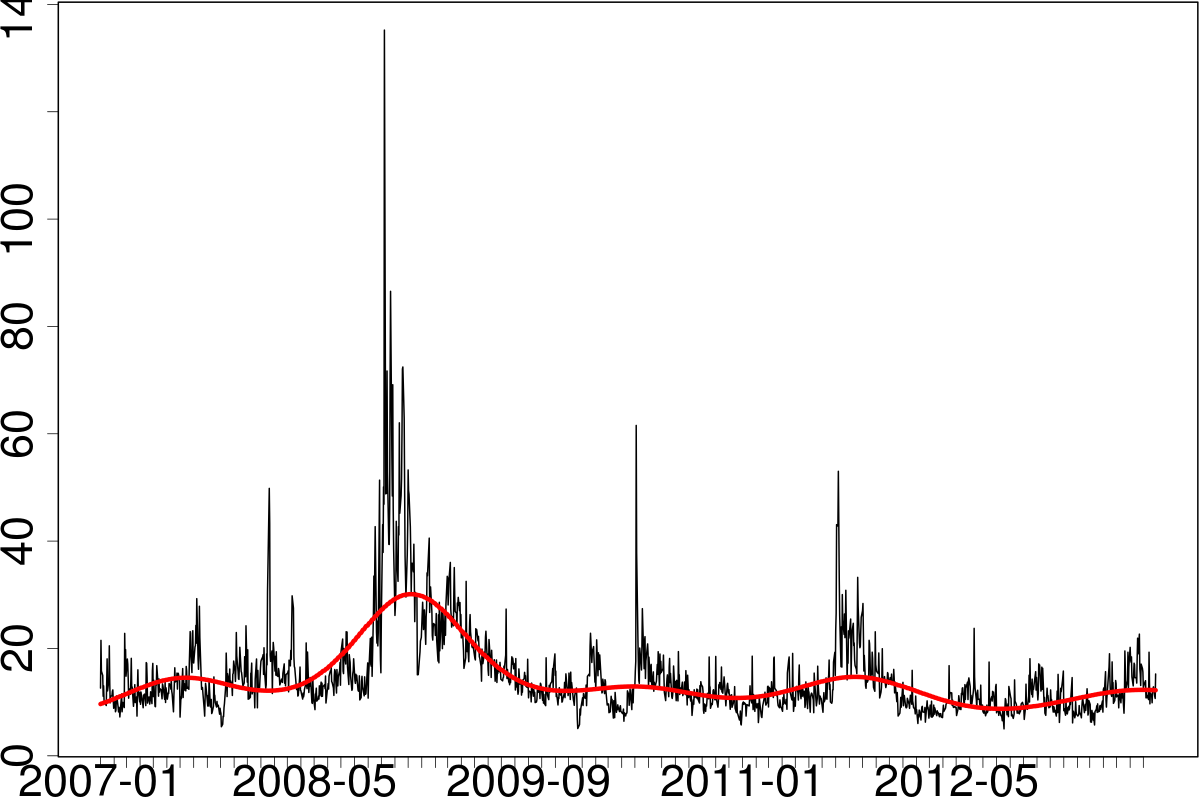}
  \end{subfigure}
  \begin{subfigure}[b]{0.49\textwidth}
      \includegraphics[width=\textwidth]{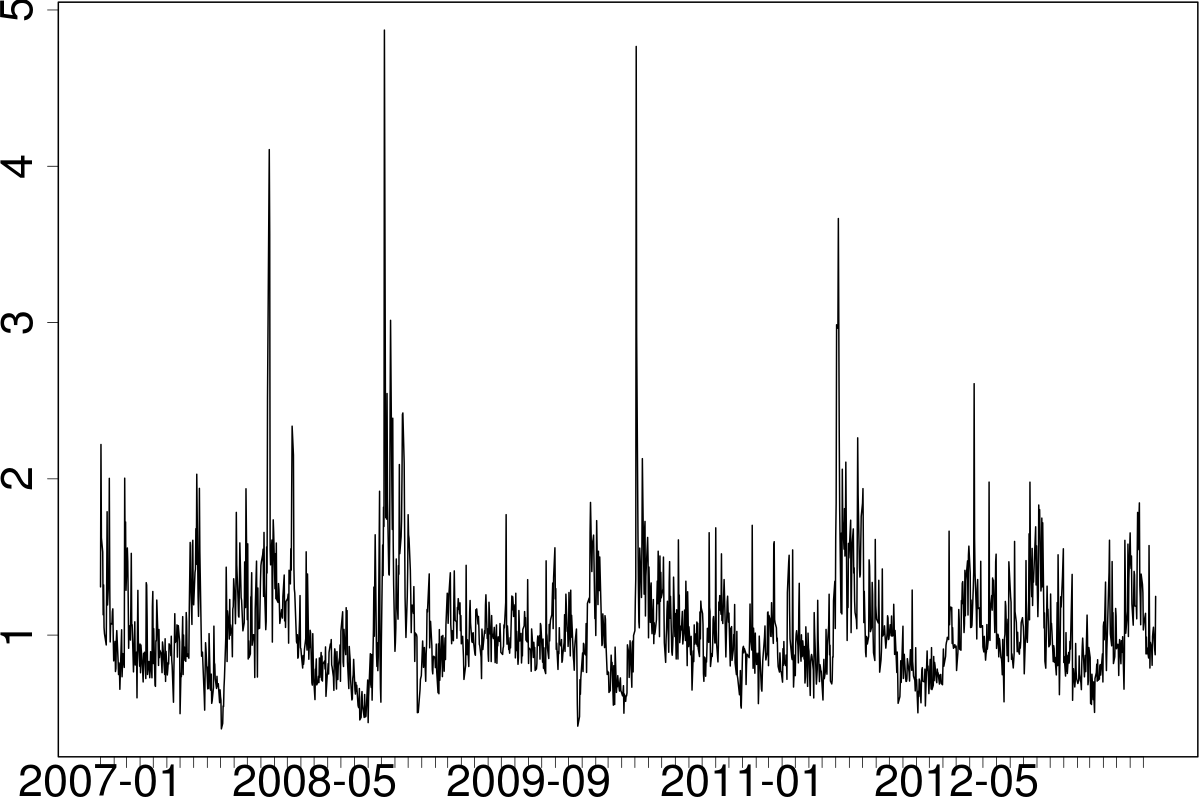}
  \end{subfigure}
  \\
  \begin{subfigure}[b]{0.49\textwidth}
      \includegraphics[width=\textwidth]{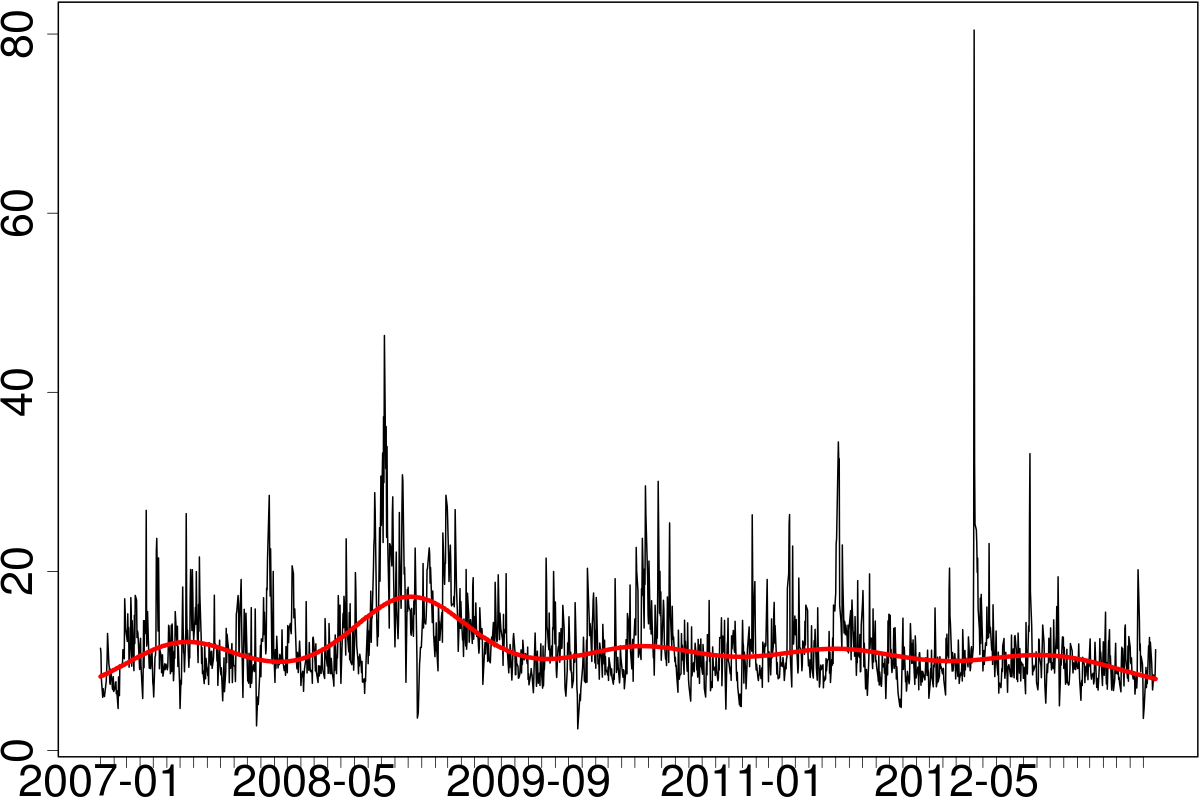}
  \end{subfigure}
  \begin{subfigure}[b]{0.49\textwidth}
      \includegraphics[width=\textwidth]{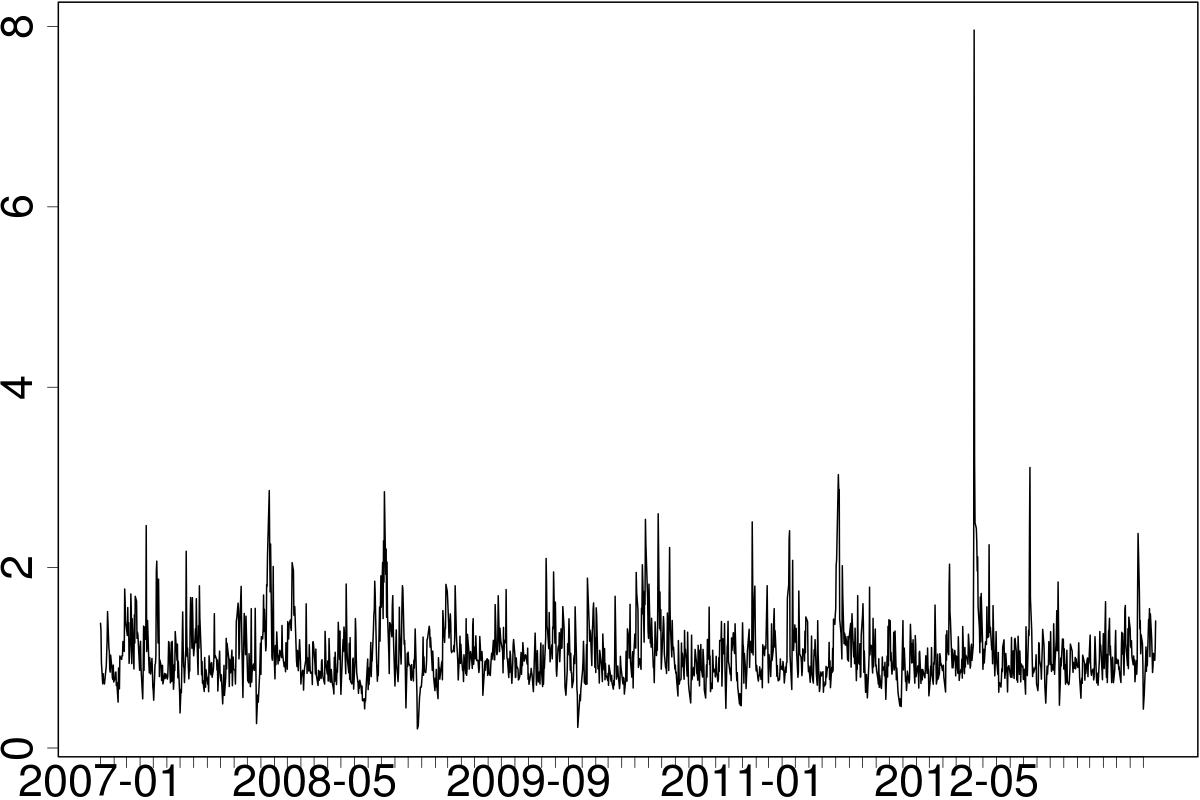}
  \end{subfigure}
  \\
  \begin{subfigure}[b]{0.49\textwidth}
      \includegraphics[width=\textwidth]{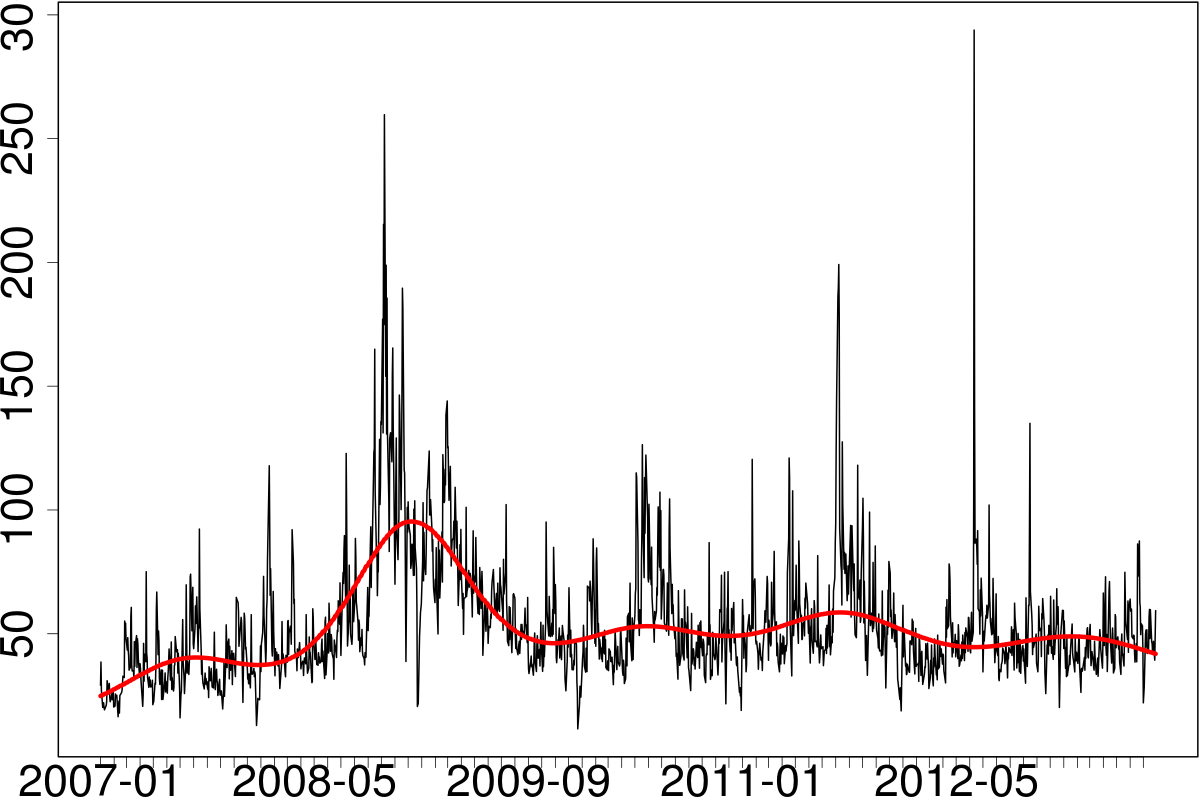}
  \end{subfigure}
  \begin{subfigure}[b]{0.49\textwidth}
      \includegraphics[width=\textwidth]{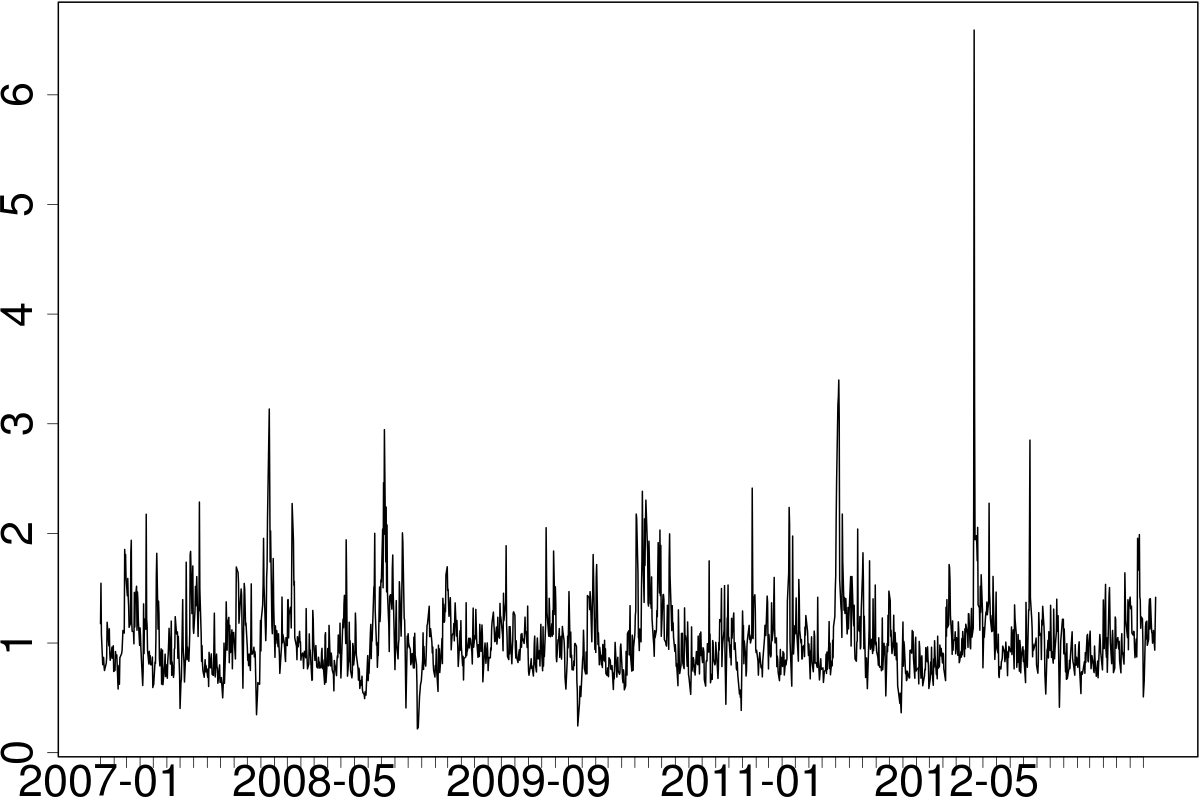}
  \end{subfigure}
\end{figure}

According to Figure~\ref{fig:JNJ-series}, the turmoil originating with the subprime mortgage crisis is clearly affecting the profile of the series with an underlying trend.
For volatility, the presence of a changing average level was analyzed by~\citet{Gallo:Otranto:2015} with a number of non-linear MEM's.
Without adopting their approach, in what follows we implement a strategy of trend removal (separately for each indicator) in order to identify short term interactions among series apart from lower frequency movements.
The presence of an upward slope in the first portion of the sample is apparent and is in line with the evidence produced by~\citet{Andersen:1996} for volumes.
Remarkably, this upward trend is interrupted after the peak of the crisis in October 2008, with a substantial and progressive reduction of the average level of the variables.
To remove the trend, we adopt a solution similar to \citet{Andersen:1996}, that is, a flexible function of time which smooths out the series.
In detail, assuming that the trend is multiplicative, we remove it in each indicator as follows:
\begin{itemize}
  \item we take the log of the original series;
  \item we fit on each log-series a spline based regression with additive errors, using time $t$ (a progressive counter from the first to the last observation in the sample) as an independent variable;%
  \footnote{Alternative methods, such as a moving average of fixed length (centered or uncentered), can be used but in practice they deliver very similar results and will not be discussed in detail here.
  The spline regression is estimated with the \texttt{gam()} function in the \texttt{R} package \texttt{mgcv} by using default settings.}
  \item the residuals of the previous regression are then exponentiated to get the detrended series.
\end{itemize}

When used to produce out-of-sample forecasts of the original quantities, the described approach is applied assuming that the trend component remains constant at the last in-sample estimate.
This strategy is simple to implement and fairly reliable for forecasting at moderate horizons.

Extracting low frequency movements in a financial market activity series with a spline is reminiscent of the stream of literature initiated by \citet{Engle:Rangel:2008} with a spline-GARCH and carried out in \citet{Brownlees:Gallo:2010} within a MEM context.
Table~\ref{tab:JNJ-Correlations} shows very similar correlations between the estimated trends, all around $0.87$.
Although quite large, correlations among the detrended series are less homogeneous: as expected, the value concerning volumes and the number of trades, above $0.9$, is the highest one; $cor(rkv, vol)$  is below $0.6$, whereas the  $cor(rkv, nt)$ is around $0.7$, confirming the intuition that the enlargement of the variables involved is warranted by the data. Note also that removing trends  tends to widen the differences among correlations when  compared with the original series.

\begin{table}[ht]
  \caption{Correlations for JNJ (Jan.~3, 2007 -- July 31, 2013).
  $rkv = $ realized kernel volatility; $vol = $ volume; $ntr = $ number of trades.}
  \label{tab:JNJ-Correlations}
  \begin{center}
  \input{Correlations.tex}
  \end{center}
\end{table}

\subsection{Modeling Results}
\label{sect:EstimResults}

In the application, we consider a vMEM on detrended data, where the conditional expectation has the form (cf.~Equation~(\ref{eqn:muVectorL}))
\begin{equation}
  \label{eqn:muVector3}
  \bm{\mu}_t = \bm{\omega} + \bm{\alpha}_1 \bm{x}_{t-1} + \bm{\alpha}_2 \bm{x}_{t-2} + \bm{\gamma}_1 \bm{x}_{t-1}^{(-)} +  \bm{\beta}_1 \bm{\mu}_{t-1}.
\end{equation}
In order to appreciate the contribution of the different model components, in this conditional mean we consider alternative specifications for the coefficient matrices $\bm{\alpha}_{1}$ and $\bm{\beta}_{1}$ ($\bm{\alpha}_2$ and $\bm{\gamma}_1$ are kept diagonal in all specifications), and for the error term.
As of the former, we consider formulations with both $\bm{\alpha}_{1}$ and $\bm{\beta}_{1}$ diagonal (labeled $D$); $\bm{\alpha}_{1}$ full and $\bm{\beta}_{1}$ diagonal (labeled $A$); both $\bm{\alpha}_{1}$ and $\bm{\beta}_{1}$ full (labeled $AB$).
For the joint distribution of the errors, we adopt a Student--T, a Normal and an Independent copula ($T$, $N$ and $I$ as respective labels), in all cases with Gamma distributed marginals.
The estimated specifications are summarized in Table~\ref{tab:JNJ-Formulations}.
When coupled with the conditional means in the table, the specifications with the Independent copula can be estimated equation--by--equation.
\begin{table}[ht]
  \begin{center}
  \caption{Estimated specifications of the vMEM defined by (\ref{eqn:vMEM_mu_eps}), (\ref{eqn:muVector3}) and (\ref{eqn:pdfEpsCopula}) with Gamma marginals.}
  \label{tab:JNJ-Formulations}
  \input{Formulations.tex}
  \end{center}
\end{table}

Estimation results are reported in Table~\ref{tab:Estimates-mu}, limiting ourselves to the equation for the realized volatility.%
\footnote{All models are estimated using Expectation Targeting (Section~\ref{sect:ExpectationTargeting}).
The Normal copula based specifications are estimated resorting to the concentrated log-likelihood approach (Section~\ref{sect:ConcentratedLogLik}).
We omit estimates of the constant term $\omega$.}
The Student-T copula turns out to be the favorite specification, judging upon a significantly higher log--likelihood function value, and lower information criteria;%
\footnote{The estimated degrees of freedom are $9.10$ (s.e.~$1.24$) and $8.72$ (s.e.~$1.13$), respectively, in the A-T and AB-T formulations.
We also tried full ML estimation of the AB-N specification getting a value of the log-likelihood equal to $2093.52$, very close to the concentrated log-likelihood approach (Section \ref{sect:ConcentratedLogLik}) used in Table~\ref{tab:Estimates-mu}.}
the equation--by--equation approach (Independent copula) is dominated in both respects. Without reporting the parameter details, an estimation of the Diagonal model with the Normal/Student-T copula function shows log-likelihood values of $1993.77$, respectively, $2076.51$ pointing to both a substantial improvement coming from the contemporaneous correlation of the innovations and to the joint significance of the other indicators when the $A$ specification is adopted.
It is interesting to note that residual autocorrelation is substantially reduced only in the case of richer parameterizations ($AB$), where both non--diagonal $\bm{\alpha}_{1}$ and $\bm{\beta}_{1}$ are allowed to capture possible interdependencies.

The impact of volumes and trades on the realized volatility is present but generally individually not significant, probably due to collinearity (as noted, a log--likelihood test would reject the joint hypothesis of the relevant coefficients being equal to zero). In general, it seems that the contribution of the number of trades is more discernible in the presence of a full $\bm{\beta}_{1}$; such a significance is also highlighted by the causality tests. The coefficients at lag $2$ are always significant with negative signs.
The Normal and the Student-T specifications appear to provide similar point estimates, except for the non-diagonal $\beta$ coefficients: for the latter, once again, the picture is clouded by collinearity, since a formal log--likelihood test does indicate joint significance.

The overview on the results is complete by examining the Table~\ref{tab:gammaM} where we report the estimated coefficient $\widehat{\phi}$ for all three Gamma marginals, showing that the estimated unconditional distributions of the estimated residuals have slightly different shapes.

\begin{table}[ht]
  \begin{center}
  \caption{Estimated coefficients of the realized volatility equation for different model formulations (cf.~Table~\ref{tab:JNJ-Formulations}) for JNJ (Jan.~3, 2007 -- July 31, 2013).
  Robust t-stats in parentheses.
  An empty space indicates that the specification did not include the corresponding coefficient.
  Causality tests (rows with the arrows) report p-values for the hypothesis $H_{0}: \alpha_{1,j} = \beta_{1,j} = 0$ ($j = 2, 3$).
  Diagnostics report Log-likelihood values, Akaike and Bayesian Information Criteria, and p-values of a joint Ljung--Box test of no autocorrelation at various lags.}
  \label{tab:Estimates-mu}
  \input{Estimates-mu.tex}
  \end{center}
\end{table}

\clearpage

\begin{table}[ht]
\begin{center}
\caption{Estimated $\phi$ parameters of the Gamma marginal distributions.} \label{tab:gammaM}
\input{Estimates-phi.tex}
\end{center}
\end{table}

Finally, the correlation coefficients implied by the copula--based specifications are reported in Table~\ref{tab:copulacorr}, showing that a strong correlation among innovations further supports the need for taking simultaneity into account.

\begin{table}[ht]
  \begin{center}
  \caption{Estimated correlation matrices of the copula functions.}\label{tab:copulacorr}
  \input{Estimates-cor.tex}
  \end{center}
\end{table}

\subsection{Forecasting}
\label{sect:Forecasting}

We left the period August 1 -- December 31, 2013 (106 observations) for out--of--sample forecasting comparisons.
We adopt a~\citet{Diebold:Mariano:1995} test statistic for superior predictive ability using the error measures
\begin{equation}
  \label{eqn:losses}
  e_{N, t} = \frac{1}{2} (x_{t} - \mu_{t})^{2}
  \qquad{}
  e_{G, t} = \ln \frac{x_{t}}{\mu_{t}} - \frac{x_{t}}{\mu_{t}} - 1.
\end{equation}
where $x_{t}$ and $\mu_{t}$ denote here the observed and the predicted values, respectively.
$e_{N, t}$ is the squared error, and can be interpreted as the loss behind an $x_{t}$ Normally distributed with mean $\mu_{t}$;
similarly, $e_{G, t}$ can be interpreted as the loss we can derive considering $x_{t}$ as Gamma distributed with mean $\mu_{t}$ and variance proportional to $\mu_{t}^{2}$.

Table~\ref{tab:DM} reports the values of the Diebold-Mariano test statistic of different model formulations, against the D-I specification, considering one-step ahead predictions.
We notice a progressive and significant improvement of the specifications allowing for more interdependencies, for both the detrended and the original series,%
\footnote{One-step ahead predictions at time $t$ for the original series are computed multiplying the corresponding forecast of the detrended indicator by the value of the trend at $t-1$.}
when the Normal-based error is considered.
As far as the Gamma-based error, significant statistics emerge for the specifications with the Student-T copula, although also those involving the Normal copula are borderline.

\begin{table}
  \begin{center}
  \caption{Diebold-Mariano test statistics for unidirectional comparisons, against the D-I formulation, considering 1-step ahead forecasts (out--of--sample period August 1 -- December 31, 2013).
  The error measures are defined as $e_{N, t} = 0.5 (x_{t} - \mu_{t})^{2}$ and $e_{G, t} = \ln(x_{t} / \mu_{t}) - x_{t}/\mu_{t} - 1$, where $x_{t}$ and $\mu_{t}$ denote the observed and the predicted values, respectively (cf.~Section~\ref{sect:Forecasting} for the interpretation).
  Boldface indicates 5\% significant statistics.}
  \label{tab:DM}
  \input{DMTest.tex}
  \end{center}
\end{table}


\section{Conclusions}

In this paper we have presented a general discussion of the vector specification of the Multiplicative Error Model introduced by~\citet{Engle:2002}: a positive valued process is seen as the product of a scale factor which follows a GARCH type specification and a unit mean innovation process.
\citet{Engle:Gallo:2006} estimate a system version of the MEM by adopting a dynamically interdependent specification for the scale factors (each variable enters other variables' specifications with a lag) but keeping a diagonal variance--covariance matrix for the Gamma--distributed innovations.
The extension to a truly multivariate process requires a complete treatment of the interdependence among the innovation terms; in this respect, the specification using multivariate Gamma distributions is too restrictive because of their limitations. One possibility is to avoid the specification of the distribution and adopt a semiparametric GMM approach as in~\citet{Cipollini:Engle:Gallo:2013}. Alternatively, and it is the avenue pursued here, we can derive a maximum likelihood estimator by framing the innovation vector as a copula function linking Gamma marginals.

We illustrate the procedure on three indicators related to market activity: realized volatility, volumes, and number of trades. The empirical results are presented in reference to daily data on the Johnson and Johnson (JNJ) stock. The data on the three variables show  a (slowly moving) time varying local average which can be removed before the rest of the analysis is performed. The three trends are highly correlated with one another, but interestingly, their removal does not have a substantial impact on the correlation among the detrended series. The specifications adopted start from the consideration of a diagonal structure where no dynamic interaction is allowed and an Independent copula (\textit{de facto} an equation--by--equation specification) as a benchmark. Refinements are obtained by inserting a Normal copula and a Student--T copula (which dramatically improve the estimated log--likelihood function values) and then allowing for the presence of dynamic interdependence. Although hindered by the presence of collinearity, the results clearly show a significant improvement for the fit of the equation for realized volatility when volumes and number of trades are considered. This is highlighted by significantly better log--likelihood, better information criteria and improved autocorrelation diagnostics. The results of an out--of--sample forecasting exercise confirm the superiority of richer specifications and a slight preference for the Student-T copula. From a substantive point of view, we interpret the results as showing that the past of realized volatility by itself is not enough information to reconstruct the dynamics and establish its forecastability as such a variable is influenced by other indicators of market activity.



\input{vMEM-arXiv.bbl}
\newpage
\appendix

\section*{Appendix: Expectation Targeting}
\label{app:exptarg}

We show how to obtain the asymptotic distribution of the estimator of the parameters of the vMEM when the constant $\bm{\omega}$ model is reparameterized, via \textit{expectation targeting} (Section \ref{sect:ExpectationTargeting}),  by exploiting the assumption of weak-stationarity of the process (Section \ref{sect:MEM}).
Under assumptions detailed in what follows, the distribution of $\overline{\bm{x}}_T$ can be found by extending the results in \citet{Francq:Horvath:Zakoian:2011} and \citet{Horvath:Kokozka:Zitikis:2006} to a multivariate framework and to asymmetric effects.

\textbf{Framework}. Let us assume a model defined by (\ref{eqn:vMEM_mu_eps}), (\ref{eqn:veps_general}) and (\ref{eqn:muVectorL}) where the
$\bm{x}_{t}^{(-)}$'s are associated with negative returns on the basis of assumptions detailed in Section \ref{sect:MEM}

Besides mean-stationarity, in order to get asymptotic normality also, we assume the stronger condition that $E(\bm{\mu}_t \bm{\mu}_t^\prime)$ exists (a similar condition on existence of the unconditional squared moment of the conditional variance is assumed in the cited papers).

\textbf{Auxiliary results.}
In order to simplify the exposition we introduce two quantities employed in the following, namely the \textit{zero mean residual}
\begin{equation}\label{eqn:v}
  \bm{v}_t = \bm{x}_t - \bm{\mu}_t
\end{equation}
and
\begin{equation}\label{eqn:xTilde}
  \widetilde{\bm{x}}_t = \bm{x}^{(-)}_t - \bm{x}_t / 2.
\end{equation}
Since $\bm{v}_t = \bm{\mu}_t \odot \left( \bm{\varepsilon}_t - \mathds{1} \right)$,
and $\widetilde{\bm{x}}_t = \bm{\mu}_t \odot \bm{\varepsilon}_t \odot \left( \bm{I}_t - \mathds{1} / 2 \right)$, we can easily check that
\begin{align}
  E(\bm{v}_t) & = \bm{0}                                                                  \nonumber
    \\
  V(\bm{v}_t) & = E(\bm{\mu}_t \bm{\mu}_t^\prime) \odot \bm{\Sigma}  \equiv \bm{\Sigma}_v \nonumber
    \\
  C(\bm{v}_s, \bm{v}_t) & = \bm{0}   \qquad s \neq t                                      \nonumber
    \\
  E(\widetilde{\bm{x}}_t) & = \bm{0}                                                      \nonumber
    \\
  V(\widetilde{\bm{x}}_t) & = \bm{\Sigma}_v \odot \bm{\Sigma}_I                           \nonumber
    \\
  C(\widetilde{\bm{x}}_s, \widetilde{\bm{x}}_t) & = \bm{0}   \qquad s \neq t              \nonumber
    \\
  C(\bm{v}_s, \widetilde{\bm{x}}_t) & = \bm{0}   \qquad s, t.                             \nonumber
\end{align}
We remark that $\bm{\Sigma}_v$ represents also the unconditional average of $\bm{x}_t \bm{x}_t^\prime$.
By consequence, the sample averages $\overline{\bm{v}}_T = T^{-1} \sum_{t = 1}^{T} \bm{v}_t$ and $\overline{\widetilde{\bm{x}}}_T = T^{-1} \sum_{t = 1}^{T} \widetilde{\bm{x}}_t$ are such that
\begin{equation}\label{eqn:v:xTilde:asyDistr}
  \sqrt{T}
    \left( \begin{array}{c}
      \overline{\bm{v}}_T
      \\
      \overline{\widetilde{\bm{x}}}_T
    \end{array} \right)
  \dlim
  N \left[
    \left(
    \begin{array}{c}
      \bm{0}
      \\
      \bm{0}
    \end{array}
    \right),
    \left(
    \begin{array}{cc}
      \bm{\Sigma}_v & \bm{0}
      \\
      \bm{0}        & \bm{\Sigma}_v \odot \bm{\Sigma}_I
    \end{array}
    \right)
    \right].
\end{equation}

\textbf{The asymptotic distribution of $\overline{\bm{x}}_T$.}

By replacing (\ref{eqn:v}) and (\ref{eqn:xTilde}) into equation (\ref{eqn:muVectorL}) and arranging it we get
\begin{equation*}
	\bm{x}_t
	  -
	\sum_{l=1}^{L}
	  \left( \bm{\alpha}_l + \bm{\beta}_l + \frac{\bm{\gamma}_l}{2} \right) \bm{x}_{t-l}
    =
  \bm{\omega}
	  +
	\bm{v}_t
	  -
	\sum_{l=1}^{L} \bm{\beta}_l \bm{v}_{t-l}
	  +
	\sum_{l=1}^{L}
	  \bm{\gamma}_l \widetilde{\bm{x}}_{t-l}^{(-)}
	\end{equation*}
so that, averaging both sides,
\begin{equation*}
	\left[
	  \bm{I} - \sum_{l=1}^{L} \left( \bm{\alpha}_l + \bm{\beta}_l + \frac{\bm{\gamma}_l}{2} \right)
	\right]
	\overline{\bm{x}}_T
    =
  \bm{\omega}
	  +
	\left[
	  \bm{I} - \sum_{l=1}^{L} \bm{\beta}_l
	\right]
	\overline{\bm{v}}_T
	  +
	\sum_{l=1}^{L} \bm{\gamma}_l \overline{\widetilde{\bm{x}}}_{T}
	  +
	O_p(T^{-1}).
\end{equation*}
Deriving $\overline{\bm{x}}_T$ we get
\begin{align*}
	\overline{\bm{x}}_T
	  & =
	\bm{\mu}
	  +
	\left[
	  \bm{I} - \sum_{l=1}^{L} \left( \bm{\alpha}_l + \bm{\beta}_l + \frac{\bm{\gamma}_l}{2} \right)
	\right]^{-1}
	\left[
	  \left( \bm{I} - \sum_{l=1}^{L} \bm{\beta}_l \right) \overline{\bm{v}}_T
	    +
	  \sum_{l=1}^{L} \bm{\gamma}_l \overline{\widetilde{\bm{x}}}_{T}
	\right]
	  +
	O_p(T^{-1}) \\
	  & =
	\bm{\mu}
	  +
	\bm{A}^{-1}
	\left[ \bm{B} \overline{\bm{v}}_T + \bm{C} \overline{\widetilde{\bm{x}}}_{T} \right]
	  +
	O_p(T^{-1}),
\end{align*}
where $\bm{\mu}$ is given in (\ref{eqn:mu}).

By means of (\ref{eqn:v:xTilde:asyDistr}), the asymptotic distribution of $\overline{\bm{x}}_T$ follows immediately as
\begin{equation}\label{eqn:xMean:asyDistr}
  \sqrt{T} \left( \overline{\bm{x}}_T - \bm{\mu} \right)
  \dlim
  N \left[
    \bm{0},
    \bm{A}^{-1}
     \left( \bm{B} \bm{\Sigma}_v \bm{B}^\prime  + \bm{C} \left( \bm{\Sigma}_v \odot \bm{\Sigma}_I \right) \bm{C}^\prime  \right)
    \bm{A}^{-1 \prime}
    \right]
\end{equation}

\end{document}

%% file: Correlations.tex
\begin{tabular}{c|cc|cc|cc|}
         & \multicolumn{2}{c}{Original} & \multicolumn{2}{c}{Trend} & \multicolumn{2}{c|}{Detrended} \\
         & $vol$     & $ntr$            & $vol$     & $ntr$         & $vol$       & $ntr$          \\ \hline
$rkv$    & 0.645     & 0.779            & 0.875     &   0.877       & 0.579       &  0.707         \\
$vol$    &           & 0.903            &           &   0.874       &             &  0.936         \\ \hline
\end{tabular}

%% file: Formulations.tex
\begin{tabular}{l|ccc|}
                                                                                               & \multicolumn{3}{|c|}{Error Distribution (copula)}  \\
\multicolumn{1}{c|}{Conditional Mean (parameters)}                                             & $I$: Independent & $N$: Normal & $T$: Student-T \\ \hline
$D$:  $\bm{\alpha}_{1}$, $\bm{\beta}_{1}$, $\bm{\gamma}_{1}$, $\bm{\alpha}_{1}$ diagonal       & D-I &      &      \\ 
$A$:  $\bm{\alpha}_{1}$ full; $\bm{\beta}_{1}$, $\bm{\gamma}_{1}$, $\bm{\alpha}_{1}$ diagonal  & A-I & A-N  & A-T  \\  
$AB$: $\bm{\alpha}_{1}$, $\bm{\beta}_{1}$ full; $\bm{\gamma}_{1}$, $\bm{\alpha}_{1}$ diagonal  &     & AB-N & AB-T \\ \hline
\end{tabular}

%% file: Estimates-mu.tex
%
\begin{tabular}{ccccccc}
 & D-I & A-I & A-N & A-T & AB-N & AB-T \\ 
  \hline
\multirow{2}{*}{$rkv_{t-1}$} & $0.4929$ & $0.4601$ & $0.4263$ & $0.4239$ & $0.3718$ & $0.3757$ \\ 
   & ($\mathit{58.78}$) & ($\mathit{53.00}$) & ($\mathit{50.39}$) & ($\mathit{50.30}$) & ($\mathit{39.53}$) & ($\mathit{41.37}$) \\ 
  \multirow{2}{*}{$vol_{t-1}$} &  & $-0.0244$ & $-0.0281$ & $-0.0241$ & $-0.0209$ & $-0.0305$ \\ 
   &  & ($\mathit{-0.89}$) & ($\mathit{-0.83}$) & ($\mathit{-0.68}$) & ($\mathit{-0.49}$) & ($\mathit{-0.72}$) \\ 
  \multirow{2}{*}{$nt_{t-1}$} &  & $0.0568$ & $0.0615$ & $0.0533$ & $0.1846$ & $0.1821$ \\ 
   &  & ($\mathit{1.53}$) & ($\mathit{1.37}$) & ($\mathit{1.22}$) & ($\mathit{2.70}$) & ($\mathit{2.74}$) \\ 
  \multirow{2}{*}{$rkv_{t-2}$} & $-0.2647$ & $-0.2453$ & $-0.1961$ & $-0.2007$ & $-0.1785$ & $-0.1843$ \\ 
   & ($\mathit{-5.16}$) & ($\mathit{-3.91}$) & ($\mathit{-3.19}$) & ($\mathit{-3.70}$) & ($\mathit{-3.84}$) & ($\mathit{-4.40}$) \\ 
  \multirow{2}{*}{$rkv_{t-1}^{(-)}$} & $0.0265$ & $0.0276$ & $0.0272$ & $0.0304$ & $0.0182$ & $0.0219$ \\ 
   & ($\mathit{0.84}$) & ($\mathit{0.83}$) & ($\mathit{0.87}$) & ($\mathit{1.00}$) & ($\mathit{0.44}$) & ($\mathit{0.54}$) \\ 
  \multirow{2}{*}{$\mu^{(rkv)}_{t-1}$} & $0.7172$ & $0.6990$ & $0.6821$ & $0.6942$ & $0.7533$ & $0.7638$ \\ 
   & ($\mathit{13.94}$) & ($\mathit{9.96}$) & ($\mathit{10.03}$) & ($\mathit{11.75}$) & ($\mathit{14.18}$) & ($\mathit{16.49}$) \\ 
  \multirow{2}{*}{$\mu^{(vol)}_{t-1}$} &  &  &  &  & $-0.1233$ & $-0.0586$ \\ 
   &  &  &  &  & ($\mathit{-0.91}$) & ($\mathit{-0.49}$) \\ 
  \multirow{2}{*}{$\mu^{(nt)}_{t-1}$} &  &  &  &  & $-0.0918$ & $-0.1463$ \\ 
   &  &  &  &  & ($\mathit{-0.64}$) & ($\mathit{-1.11}$) \\ 
   \hline
  $rkv_{t} \leftarrow vol_{t-1}$ &  & 0.3733 & 0.4083 & 0.4979 & 0.2530 & 0.3187 \\ 
  $rkv_{t} \leftarrow nt_{t-1}$ &  & 0.1268 & 0.1703 & 0.2224 & 0.0124 & 0.0133 \\ 
   \hline
logLik & 238.24 & 270.87 & 2062.12 & 2133.07 & 2092.77 & 2168.15 \\ 
  AIC & -440.48 & -493.73 & -4070.25 & -4210.14 & -4119.53 & -4268.30 \\ 
  BIC & -323.28 & -337.47 & -3894.46 & -4027.83 & -3904.68 & -4046.94 \\ 
   \hline
LB(12) & 0.0000 & 0.0000 & 0.0005 & 0.0002 & 0.0552 & 0.0239 \\ 
  LB(22) & 0.0000 & 0.0000 & 0.0003 & 0.0002 & 0.0291 & 0.0195 \\ 
  LB(32) & 0.0000 & 0.0001 & 0.0015 & 0.0012 & 0.0357 & 0.0261 \\ 
   \hline
\end{tabular}

%% file: Estimates-phi.tex
\begin{tabular}{ccccccc}
           & D-I & A-I & A-N & A-T & AB-N & AB-T \\ 
  \hline
$\phi_{1}$ & 23.08 & 23.22 & 23.31 & 22.34 & 23.70 & 22.62 \\ 
  $\phi_{2}$ & 15.56 & 16.01 & 15.82 & 15.38 & 15.89 & 15.46 \\ 
  $\phi_{3}$ & 19.55 & 19.63 & 19.23 & 18.23 & 19.34 & 18.36 \\ 
  \end{tabular}

%% file: Estimates-cor.tex
\begin{tabular}{c|cccccccc}
   & \multicolumn{2}{c}{A-N} & \multicolumn{2}{c}{A-T} &  \multicolumn{2}{c}{AB-N} & \multicolumn{2}{c}{AB-T} \\ 
   & $vol_{t}$ & $nt_{t}$ & $vol_{t}$ & $nt_{t}$ & $vol_{t}$ & $nt_{t}$ & $vol_{t}$ & $nt_{t}$ \\ 
   \hline
$rkv_{t}$ & 0.479 & 0.606 & 0.499 & 0.621 & 0.481 & 0.609 & 0.502 & 0.625 \\ 
  $vol_{t}$ &  & 0.902 &  & 0.917 &  & 0.903 &  & 0.917 \\ 
  \end{tabular}

%% file: DMTest.tex
\begin{tabular}{c|cccc}
              & \multicolumn{2}{c}{$e_{N,t}$}        & \multicolumn{2}{c}{$e_{G,t}$}    \\	
	Formulation & detrended	      & original           &	detrended      & original       \\ \hline
  A-I	        & 0.924	          & 0.949	             &  1.043	         & 1.050          \\
  A-N	        & 1.644         	& \textbf{1.754} 	   &  1.564          & 1.571          \\
  A-T	        & \textbf{1.821}  & \textbf{1.950} 	   &  \textbf{1.763} & \textbf{1.768} \\
  AB-N	      & \textbf{2.004}	& \textbf{1.940} 	   &  1.641    	     & 1.649          \\
  AB-T	      & \textbf{2.128}	& \textbf{2.088}	   &  \textbf{1.807} & \textbf{1.814} \\
\end{tabular}